\documentclass[useAMS,usenatbib]{mn2e}
\usepackage{times}
\usepackage{graphicx}
\usepackage{natbib}
\usepackage{multirow}

\newcommand{\tab}{\hspace*{2em}}

\title[Spectral Properties \& z Cut-off of Compact AGNs]{Spectral
Properties and the Effect on Redshift Cut-off of Compact AGNs from the AT20G Survey}%\\
\author[R. Chhetri et
al.] {R. ~Chhetri$^{1,2}$\thanks{email:rchhetri@phys.unsw.edu.au},
R. D. ~Ekers $^{2}$, E. K. ~Mahony$^{2,3}$, P. A. ~Jones$^{1}$, M.
~Massardi$^{4,5}$,
\newauthor 
R. ~Ricci$^{4}$ and E. M. ~Sadler$^{3}$ \\
$^{1}$Department of Astrophysics \& Optics, School of Physics, University of
New South Wales, NSW 2052, Australia\\
$^{2}$Australia Telescope National Facility, CSIRO Astronomy and Space Science,
P.O. Box 76, Epping, NSW
1710, Australia\\
$^{3}$Sydney Institute for Astronomy, School of Physics, The University of
Sydney, NSW 2006, Australia\\
$^{4}$Instituto di Radioastronomia, INAF, Via Gobetti 101,
40129 Bologna,
Italy\\
$^{5}$INAF - Osservatorio Astronomico di Padova, Vicolo dell'Osservatorio 5,
I-35122 Padova, Italy}

\begin{document}

\maketitle
%\linenumbers

\begin{abstract}
We use high angular resolution data, measured from visibility of sources at the longest baseline of 4500 m of the Australia Telescope Compact Array (ATCA), for the Australia Telescope 20 GHz (AT20G) survey to obtain angular size information for $>$ 94\% of AT20G sources. We confirm the previous AT20G result that due to the high survey frequency of 20 GHz, the source population is strongly dominated by compact sources (79\%). At 0.15 arcseconds angular resolution limit, we show a very strong correlation between the compact and extended sources with flat and steep-spectrum sources respectively. Thus, we provide a firm physical basis for the traditional spectral classification into flat and steep-spectrum sources to separate compact and extended sources. We find the cut-off of -0.46 to be optimum for spectral indices between 1 and 5 GHz and, hence, recommend the continued use of -0.5 for future studies.

We study the effect of spectral curvature on redshift cut-off of compact AGNs using recently published redshift data. Using spectral indices at different frequencies, we correct for the redshift effect and produce restframe frequency spectra for compact sources for redshift up to $\sim$ 5. We show that the flat spectra of most compact sources start to steepen at $\sim$ 30 GHz. At higher frequencies, the spectra of both populations are steep so the use of spectral index does not separate the compact and extended source populations as well as in lower frequencies. We find that due to the spectral steepening, surveys of compact sources at higher frequencies ($>$ 5 GHz) will have redshift cut-off due to spectral curvature but at lower frequencies, the surveys are not significantly affected by spectral curvature, thus, the evidence for a strong redshift cut-off in AGNs found in lower frequency surveys is a real cut-off and not a result of K-correction.

\end{abstract}

\section{Introduction}
Strong evolution of the number density of Active Galactic Nuclei (AGNs) with
redshift has been
discussed in many papers. Studies done in optical, X-ray and radio
frequencies have  identified a ``quasar epoch", where the number density rises
sharply as a function of redshift from z = 0 to redshift of about
2.5 to 3. The space density has been shown to decline beyond the redshift of 3
in flat-spectrum radio sources
(eg. \citealt{Peacock1985, Dunlop1990, Shaver1996, Wall2005}), in X-ray
selected quasars \citep{Hasinger2005} and optically selected quasars
\citep{Schmidt1991, Vigotti2003}. The details of whether such an
evolution is purely due to space density evolution or due to luminosity
evolution is yet to be resolved. 
Similarly, the decline in the number
density in
steep-spectrum radio sources has also been discussed in the literature (eg. 
\citet{Dunlop1990, Jarvis2001}) and
has been studied in detail by \citet{Rigby2011}.

The decline in the number density of QSO population beyond redshift
$\sim$ 3 has
been termed ``redshift cut-off" and has been explored in some detail by many
authors. Some explanations offered suggest luminosity dependent space density
evolution \citep{Dunlop1990} while others suggest luminosity evolution
\citep*[e.g.][]{Boyle1988}. The rate of decline in the number density has also been
discussed in detail. For example, \citet{Dunlop1990} and \citet{Vigotti2003} suggest a slow
decline while \citet{Shaver1996} and \citet{Wall2005} find a rapid decline in
the number density of flat-spectrum radio quasars at high redshift, comparable
to results from X-ray and optical work. However, \citet{Jarvis2000} have argued
that spectral curvature causes the observable volume at high redshift universe
to be too small to find any luminous flat-spectrum sources, removing the
evidence for the rapid decline in number density at high redshift suggested by
\citet{Shaver1996}. The spectral
curvature of sources at different redshifts gives the impression of redshift
cut-off by causing the sources to fall outside of selection criteria either by
making the flux density fall under the selection limit or the spectral
index to fall outside the selection limit, in populations selected based on
their spectra (eg. flat-spectrum population). \citet{Wall2005} have
found that the use of non-contemporaneous flux density data can cause the effect
of steep spectral curvature similar to that seen by \citet{Jarvis2000} (see also
\citet{de_Zotti2010}). 

The abundance of
quasars at redshifts 2.5 - 3 has implications for the epoch of formation
of the first structures in the universe, which has important consequences for the
formation and evolution of massive galaxies. This epoch also has importance in
determining whether AGNs can be significant sources of reionization of the
universe.

Radio frequency studies of the redshift cut-off have the
advantage of being free from dust obscuration. Such studies have been done
in the past with
populations of flat-spectrum sources, selected
from low frequency ($\leq$ 5 GHz) source catalogues. Flat-spectrum sources have
traditionally been associated with compact sources \citep{Wall1977} with the spectral transition
placed variably around $-0.4$ (eg. \citealt{Shaver1995, Wall2005}),
$-0.5$ (eg. \citealt{Peacock1985, Saikia2001, Browne2003}) or $-0.6$ (eg.
\citealt{DeBreuck2000}). In this paper, we follow the
convention of $S_\nu\propto \nu^\alpha$ and use
-0.5 as the flat-spectrum cut-off. The flat-spectrum arises
from the cumulative effect of multiple, very compact, self absorbed components
peaking at different frequencies. Thus, compact AGNs are
considered to have a very flat-spectrum, often spanning decades of frequencies.

The Australia Telescope 20 GHz (AT20G) survey \citep{Murphy2010} is a complete
sample of radio
sources and is dominated by compact QSOs, due to its high frequency selection
\citep{Sadler2006}. These
sources have spectra measured at one to three different frequencies (20 GHz, 8.6
GHz and 4.8 GHz) and have counterparts at one to two low frequencies
($\sim$ 1 GHz) . The three
high frequencies were measured near simultaneously as part of the survey to
remove any
effects of variability. \citet{Sadler2006} find the median variability index at 20 GHz
to be 6.9 per cent with 5 per cent of sources varying by $>$ 30 per cent in flux density on a 1-2 year time-scale. The AT20G
sources 
also have an angular size measurement at sub arcsec resolution which cleanly
separates the compact AGNs
from their jets and extended radio-lobes. In this paper, we demonstrate
the strong correlation between compact sources and flat-spectrum sources
providing a firm physical basis for the use of spectral index for such a
separation. We then look at the radio spectrum of these compact sources to
investigate the effect of redshift cut-off.

In section 2, we outline the source sample used for this investigation. In
section 3, we explain the use of visibility to obtain an angular size
measurement. Section 4 discusses the measurement of 
spectra of compact sources. Section 5 discusses the properties of such
compact sources. In section 6 we investigate the effect of 
spectral curvature and its effect on the redshift cut-off of QSOs. Section 7 is
the summary of this work. $\Lambda$CDM cosmology with $H_0 = 71 ~\rm{ km s^{-1}
Mpc^{-1}}$, $\Omega_m = 0.27$ and $\Omega_\Lambda = 0.73$ \citep{Larson2011}
has been used for this paper.

\section{The AT20G sample}
The AT20G is a blind extragalactic survey carried out at a high frequency of 20
GHz
for all declinations of the southern sky using the Australia Telescope Compact
Array (ATCA). It covers a total area of 20, 086 deg$^2$. It has a galactic
latitude cut-off $|b|=1.5$ degrees. The survey has a total of 5890 sources
above the flux density limit of 40 mJy at 20 GHz. The AT20G is the largest blind
survey done at such a high radio frequency. Most sources south of the
declination of -15 degrees have near-simultaneous follow-up observations
within a month at 4.8 and 8.6 GHz
\citep{Murphy2010}. The sources not observed at 4.8 and 8.6 GHz were a result of
scheduling logistics and can not introduce any spectral bias. The cross-match of AT20G sources
against the 1.4 GHz NVSS 
\citep{Condon1998} catalogue for sources north of $\delta = -40$ degrees and
0.843 GHz SUMSS \citep{Mauch2003} catalogue for sources south of $\delta = -30$
degrees provides between one to two spectral points near 1 GHz resulting in flux
densities at two to five frequencies for the sample. Although the lower
frequency points
are not simultaneous, there is relatively little variability at these
frequencies. Despite the galactic cut-off, the AT20G catalogue still contains a
small number of galactic thermal sources which are extended sources of optically
thin (flat-spectrum) free-free emission. These high latitude sources are easily
seen in H$\alpha$ surveys and have been identified and removed from the main
population for this analysis of the extragalactic sources.

Interferometric visibilities on long spacings (henceforth: 6km visibility) for
$>94\%$ of the AT20G sources have been measured (see section 3). 3403 AT20G
sources have four
or more spectral points between 1 and 20 GHz as well as 6km visibility. Optical
identifications for 3873 AT20G sources have been made and redshifts of 1460
sources have been compiled \citep{Mahony2011}.
This work makes use of the subpopulation of 838 sources with measured redshift,
6km visibility and flux densities measured at four to five frequencies. Table
\ref{tab-at20g_subpop} provides a summary of the different subpopulations. 

\begin{table}
\caption{Number of AT20G sources with follow-ups, cross-matches with low
frequency catalogues and optical counterparts.}	
  \begin{normalsize}
  \begin{center}
  \begin{tabular}{ll}
  %\begin{table}
	  \hline
	  Population & Number\\
	  %\hline
	  \hline
	  Total AT20G & 5890\\
	  6km visibility measured & 5542\\
	  $AT20G^1$ & 3795 \\
	  $AT20G^1$ + 6km visibility + NVSS/SUMSS & 3403\\
	  AT20G + Optical ID & 3873\\
	  AT20G + Optical ID + Redshift & 1460\\
	  AT20G + Optical ID + Redshift +  & \\
	  \tab \tab 6km visibility + NVSS/SUMSS &1377\\
	  $AT20G^1$ + Optical ID + Redshift + & \\ 
	  \tab \tab 6km visibility + NVSS/SUMSS & 838\\
	  \hline

  \end{tabular}
  \end{center}
  \end{normalsize}
\label{tab-at20g_subpop}
\medskip
$AT20G^1$ is the subpopulation of AT20G with 8.6 and 4.8 GHz follow-up. Sources
with $\delta > -15\degr$ do not have 8.6 and 4.8 GHz follow-up.
\end{table}

\begin{figure}
	\centering
	\begin{center}
	\includegraphics[width=0.5\textwidth]{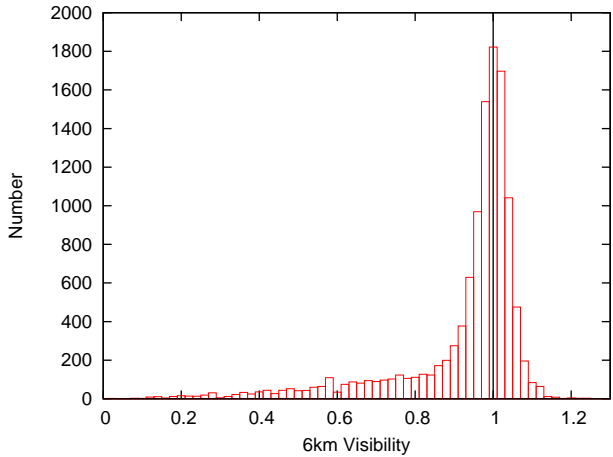}

\end{center}

	\caption{The distribution of all 6km visibilities in the
AT20G. Unresolved sources have a visibility of 1 marked by the solid
vertical line.}
	\label{fig:6kvis-histo}
\end{figure}

\section{The measurement of source size}

The AT20G follow-up was done with the hybrid configurations of the ATCA, with
five antennae in a compact configuration and the sixth antenna (called
the 6km antenna in this paper) at a baseline of $\sim 4500$m from the location
of the compact configuration.
This allowed us to calculate a visibility amplitude for over 94\% of AT20G
sources
(Table \ref{tab-at20g_subpop}). The remaining sources do not have their 6km
visibility due to inclement weather during observation or
unavailability of the 6km antenna due to maintenance. Thus, the sources with the
visibility data are free from any bias.

We calculate the 6km visibility from the ratio of the scalar average of the five
long baseline amplitudes to the scalar average of the 10 short baseline
amplitudes for the AT20G sources at 20 GHz (Chhetri et al. in prep). Although,
the AT20G follow-up observations were done in a hybrid mode,
the ATCA when used with its 6km antenna is essentially a linear array. So, the
6km visibility corresponds to a position angle on the sky which is dependent
upon the hour angle of observation. Most AT20G sources have visibilities
measured at two or more widely separated hour angles. Fig. \ref{fig:6kvis-histo}
shows the distribution of all AT20G 6km visibilities for all sources. The 6km
visibility for a point sources is expected to be 1. The Fourier transform
provides a direct relationship between visibility and angular size of a source.
Such a source size relationship is model dependent (eg. Gaussian, double
peaked etc.)
but, none the less, provides an effective method of separating compact sources
from extended sources based on their size.

Since it is not possible for any source to have visibility $>$
1, we can use the distribution of
visibilities $>$ 1 as an empirical estimate of the effect of noise on the
visibility estimate. Visibilities
found to be $>$ 1 in Figs. \ref{fig:6kvis-histo} and \ref{fig:6kvisAlphaBasic}
are due to the
effects of random noise on 6km amplitudes of unresolved sources with a RMS
scatter of 0.053 to a maximum of 1.14. Considering a symmetric noise related
distribution from the unresolved value of 1, we arrive at 0.86, where we
separate the compact sources from the extended sources and only 0.5$\%$ of
compact sources would be misclassified as extended. Therefore, for the purpose
of this paper, we define sources with 6km visibility $> 0.86$ to be compact
sources. For a Gaussian source, a visibility of 0.86 at 4500 m
corresponds to a size of 0.15 arcsecond at 20 GHz. 

We find that 79\% of the sources in our
survey are compact AGNs which is a result of the high survey frequency. This is
very different from low frequency surveys such as NVSS which has only 25\% of
the flat-spectrum AGN source population \citep{DeBreuck2000}.

\begin{figure}
	\centering
	\begin{center}
	\includegraphics[width=0.5\textwidth]{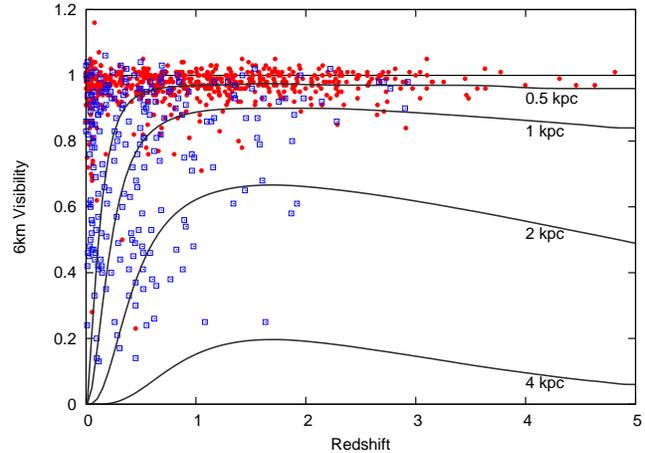}

\end{center}

	\caption{Redshift distribution of 6km visibility for flat (filled
circles, $\alpha_{1}^{4.8} > -0.5$) and steep-spectrum (open squares,
$\alpha_{1}^{4.8} \leq -0.5$) sources. Visibility models
are drawn for 0.5, 1, 2 and 4 kpc diameter Gaussian sources.}
	\label{fig:6kvis-z}
\end{figure}

In Fig. \ref{fig:6kvis-z} we have plotted the visibility on the 6km
antenna baselines against redshift. Models for sources with an assumed Gaussian
distribution of
brightness are shown for a range of linear sizes (full width at half
maximum).  If the source brightness distribution is double peaked rather than
Gaussian the curves have exactly the same form but the sizes are scaled down by
a factor
of 1.5.   In reality there is variation in source morphologies
and a single visibility cannot distinguish between these structures so the
lines in Fig. \ref{fig:6kvis-z} correspond to a small range of
possible sizes. However, this uncertainty resulting from unknown source
morphology is a relatively small fraction of the known range in radio source
sizes from cores in the central few hundred parsec to megaparsec scale lobes.
 
One interesting variation in the interpretation of the visibility occurs for
sources with emission coming from multiple regions
with very different linear sizes. For example sources which have
a significant flux in an unresolved core ($<$0.15 arcsec in our case) and the
rest
of the flux in a resolved jet or lobe structure.  In this case the visibility
should be interpreted as the fraction of the flux in the unresolved component,
and the curves in Fig. \ref{fig:6kvis-z} indicate the scale size for which the
core component is unresolved.

\section{Properties of compact sources}
\begin{figure}
	\centering
	\begin{center}
	\includegraphics[width=0.5\textwidth]{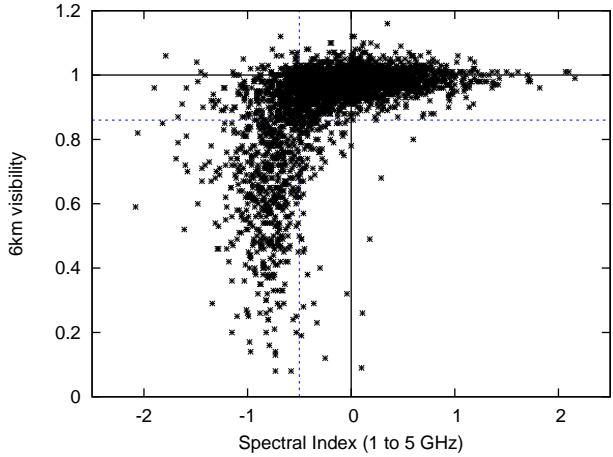}

\end{center}

	\caption{Spectral index distribution plotted against 6km visibility.
The dotted
lines mark the limits of flat v steep-spectrum on the x-axis and compact v
extended sources on the y-axis. Solid line at y = 1 represents the expected
values of visibility for unresolved sources. The flux density for 1 GHz is
obtained either from SUMSS or NVSS catalogue while 5 GHz is obtained from
AT20G follow-up at 4.8 GHz.}
	\label{fig:6kvisAlphaBasic}
\end{figure}

Over the years, the spectral index of a radio source has been the primary method
for selecting particular class of sources such as compact steep
spectrum (CSS) sources, gigahertz peaked spectrum (GPS) sources etc. (eg.
\citealt{O'Dea1998, Saikia2001}); for selecting candidates for gravitationally
lensed flat-spectrum sources (eg. \citealt{Myers1995, Winn2000, Browne2003});
for the classification of populations such as beamed and unbeamed (eg.
\citealt{Jackson1999}); for modelling of source populations (eg.
\citealt{Dunlop1990,Jarvis2000, Wall2005, de_Zotti2005, Massardi2010}); and for
physical models (eg.
\citealt{Tucci2011}). Fig. \ref{fig:6kvisAlphaBasic} plots spectral index
between 1 and 4.8 GHz against their 6km visibility for 3403 sources. With the
separation between compact and extended sources at the 6km visibility of 0.86,
there is a very clean separation between compact flat-spectrum and extended
steep-spectrum sources at the $\alpha \sim $ -0.5 line. The remarkably clean
separation of compact flat-spectrum sources and extended steep-spectrum sources
gives a very firm physical basis for the traditionally used spectral index as a
separator of compact and extended sources. Only 111 out of 2347 (5\%) flat
spectrum sources are extended while 406 out of 3160 (12\%) compact sources are
steep-spectrum sources. 

\begin{figure}
	\centering
	\begin{center}
	\includegraphics[width=0.5\textwidth]{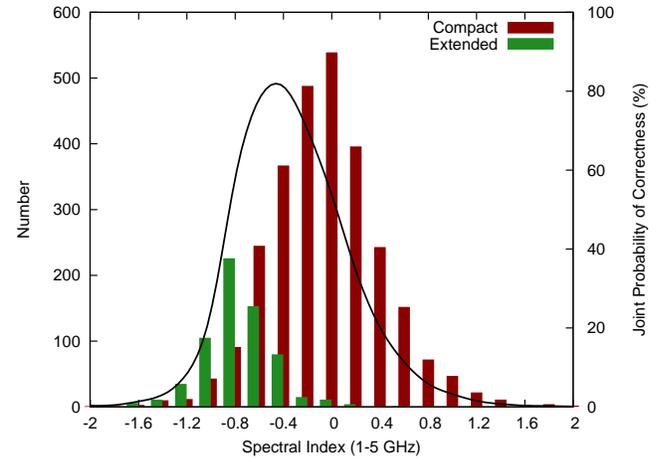}

\end{center}
	\caption{Distribution of Compact and Extended sources over spectral
index. The spectral indices are calculated between SUMSS/NVSS 1 GHz and AT20G
4.8 GHz flux densities. Mean spectral indices and dispersions from Gaussian fits
to extended and compact source population distributions are given in Table
\ref{tab:compareGaussians}. The solid curve is the joint probability of
correctness (see text). The histogram for extended sources are shifted by -0.05
to show both populations at the steep-spectrum end.}
	\label{fig:histo_alpha-6kvis}
\end{figure}

Fig. \ref{fig:histo_alpha-6kvis} shows the distribution of spectral index
($\alpha_1^{4.8}$) for both compact and extended source populations. The mean
and standard deviation for Gaussian fits for the compact and extended source
populations are given in Table \ref{tab:compareGaussians}.

We can use the
spectral index distributions for compact and extended sources
plotted in Fig. \ref{fig:histo_alpha-6kvis} to find the optimum value of the
spectral cut-off to classify the flat and steep-spectrum AGN populations.  We
get slightly different answers depending on what we want to optimise.  The
maximum value of the joint probability that both flat and steep-spectrum sources
are correctly classified is at $\alpha$ = -0.46 and 80$\%$ of the sources are
correctly classified (see Fig. \ref{fig:histo_alpha-6kvis}).  This is
independent of the relative number of flat and steep-spectrum sources in the
survey, and consequently is only weakly dependent on the survey frequency.  This
estimate is right between the two values (-0.4 and -0.5) mostly used
historically, for
consistency we recommend the use of a -0.5 cut-off in future studies.
However, if we want to maximise the total number of correctly classified
compact sources
this does depend on the relative amplitude of the two populations and is
dependent on survey frequency varying between $\alpha$ = -0.4 for a survey
frequency of 5 GHz to $\alpha$ = -0.6 for a survey frequency of 20 GHz.

\begin{table}
\caption{Gaussian fits to the spectral index distribution of compact and
extended sources in Fig. \ref{fig:histo_alpha-6kvis}.}	
  \begin{small}
  \begin{center}
  \begin{tabular}{cccc}
  %\begin{table}
	  \hline 
	  Survey Frequency	&178 MHz &\multicolumn{2}{c}{20 GHz}\\
	  \hline
	  Data 		& 3C$^{*}$	& AT20G Extended&
\textit{\textbf{AT20G Compact}}\\
	  %\hline
	  \hline
	  Mean 		& -0.85		&-0.75 &
\textit{\textbf{0.00}}\\
	  Sigma		& 0.20		&0.23 &
\textit{\textbf{0.42}}\\
	  \hline

  \end{tabular}
  \end{center}
  \end{small}
\label{tab:compareGaussians}
\medskip  *3C values for galaxies for spectral index between 0.75 and 5 GHz
from \citet{Kellermann1969}.
\end{table}

\section{The spectra of sources with 6km visibility}
\begin{figure}
	\centering
	\begin{center}
	\includegraphics[width=0.5\textwidth]{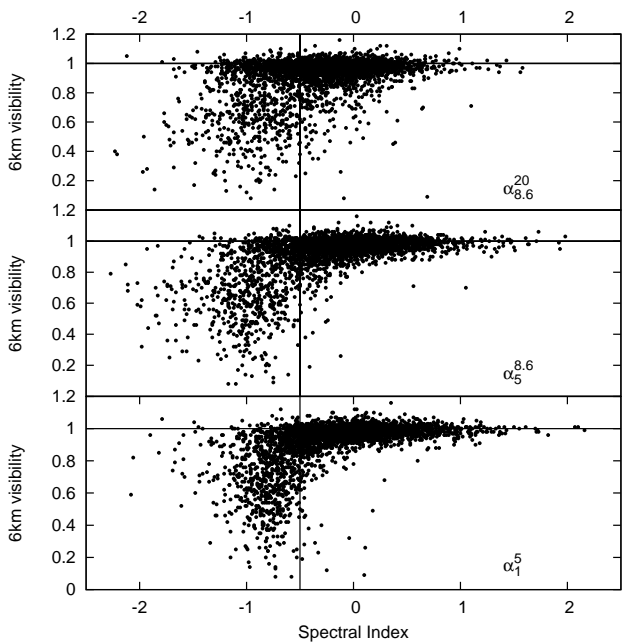}

\end{center}

	\caption{6km visibility is plotted against three spectral indices
increasing in frequencies with $\alpha_1^{4.8} $ at the bottom panel,
$\alpha_{4.8}^{8.6} $ at the mid panel and $\alpha_{8.6}^{20}$ at the top panel.
The solid vertical line corresponds to spectral index = -0.5 and the solid
horizontal lines correspond to 6km visibility = 1.}
	\label{fig:multiplotAlphaGamma}
\end{figure}

The overall spectral properties of AT20G sources show complex spectral
changes and flux density dependence \citep{Massardi2011}. In Fig.
\ref{fig:multiplotAlphaGamma} we compare
the spectra for compact and extended sources for three different frequency
ranges. We note the different characteristics between the three frequencies.
Spectral index between two frequencies with larger
separation is less susceptible to errors due to flux measurement errors than
spectral index between two frequencies close to each other. For a 10\% error in
flux density in Fig. \ref{fig:multiplotAlphaGamma}, the $\alpha_{5}^{8.6}$ has
an error of 0.16, the $\alpha_{8.6}^{20}$ has an error of 0.11 and
$\alpha_{1}^{5}$ has an error of 0.06. The 5 and 8
GHz flux density measurements were done within a month of the 20 GHz
measurements. Hence, they do not suffer much from variability of compact
sources \citep{Massardi2011} . Although the 1 GHz flux densities were measured
at different epochs (between minimum of one to maximum of 15 years), variability
is much less at lower frequencies. Hence, errors in flux densities or due to
variability
do not explain most of the differences seen in the three plots.

In Fig. \ref{fig:multiplotAlphaGamma}, we note the
appearance of a steep-spectrum `tail' for compact sources in the top
$\alpha_{8.6}^{20}$ plot. This indicates that while compact sources are mostly
flat
spectrum
at lower frequencies, a fraction of compact sources have steeper spectra between
8 and 20 GHz. It is also evident that the population of compact inverted
spectrum sources, sources with spectral index $>$ 1, at lower
frequencies shifts to lower spectral index and combines with the main population
of the compact sources at higher frequencies. These are compact core sources
that are rising at lower frequencies and have either become flat or steep at
higher frequencies (e.g. GPS sources) \citep{Sadler2008}.  

On the other hand the spectra of extended
sources which are well separated by spectral index at low frequencies, have a
larger scatter at
higher frequencies. A significant fraction of the extended sources move
towards the flatter part of the plot at higher frequency. These extended sources
are                                        
mostly composed of a flat-spectrum core with a steep-spectrum jet, lobe or a
hotspot. With increasing frequency, the steep-spectrum part of the sources gets
weaker while the compact flat-spectrum cores start to
dominate the spectrum. These composite sources show an overall combined flat
spectrum but are still extended. This domination of flat-spectrum core should be
easily observable by imaging these composite sources at higher frequencies, when
only the compact cores should be visible. With increasing frequency, there is
a further steepening in the spectra of the extended structures due to energy
loss. This is evident in the left part of Fig.
\ref{fig:multiplotAlphaGamma} with the appearance of a population of very steep
extended sources in the $\alpha_{5}^{8.6}$ and $\alpha_{8.6}^{20}$ plots.
These two effects combined produce
the larger scatter seen in the $\alpha_{8.6}^{20}$ plot of Fig.
\ref{fig:multiplotAlphaGamma} for extended sources.  

It should also be noted that at higher frequencies the spectra of compact
sources starts to become indistinguishable from steep-spectrum sources (Fig.
\ref{fig:multiplotAlphaGamma}, $\alpha_{8.6}^{20}$ plot). It is
then clear that while the use of spectral index to separate compact and extended
sources is valid using lower frequency spectra, it is unable to provide a clean
separation of compact and extended sources at higher frequencies.

\section{The redshift cut-off and the effect of spectral curvature}
\citet{Wall2005} used 379 radio selected QSOs
from the Parkes quarter-Jansky flat-spectrum sample selected at 2.7 GHz to study
the evolution of flat-spectrum quasars at high redshift. They found that the
redshift
cut-off is real and is similar to the cut-off observed in optically selected
QSOs. Similar results have been reported for cosmic star formation rates.
However, \citet{Jarvis2000}, using the Parkes half-Jansky flat-spectrum sample
selected at 2.7 GHz, argue that because of spectral curvature and the resulting
K-correction there is not enough observable volume at high redshift to find 
the brightest flat-spectrum radio sources, which gives the impression
of redshift cut-off. We can use the
sub population of compact sources from the AT20G survey, with redshifts, to study
the effect of spectral curvature and the K-correction on the redshift cut-off.

A survey at any frequency is biased to the sources that are
stronger at the survey frequency. Since this is very obvious, it is not usually
considered a bias. At lower frequencies the steep-spectrum sources dominate the
population at all luminosities and only contain the rare extremely luminous
flat-spectrum sources. Since it is the flat spectrum population that has a high
quasar identification rate, this is the sample used to investigate evolution
of flat-spectrum quasars (\citealt{Dunlop1990, Shaver1996, Jarvis2000,
Vigotti2003, Wall2005}). Our frequency selection provides a much larger fraction
of
these sources without introducing a different class of sources, thus, we
reduce the bias by having a high frequency selected sample
(e.g. \citealt{Condon2009}).

\begin{figure}
	\centering
	\begin{center}
	\includegraphics[width=0.5\textwidth]{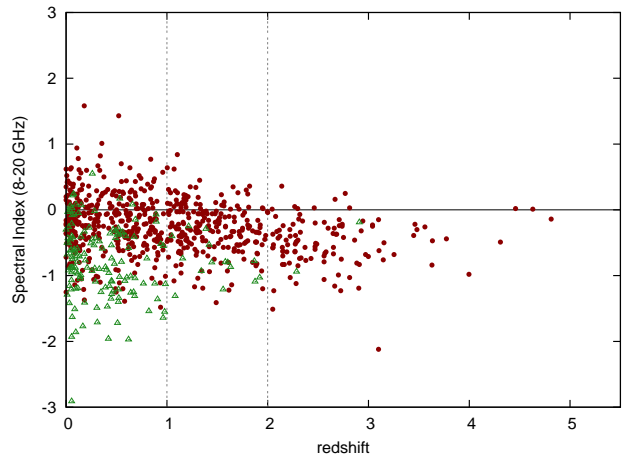}

\end{center}

	\caption{Spectra of compact and extended sources with redshift
calculated between 8.6 and 20 GHz. Filled circles represent compact sources
and open triangles represent extended sources. The median and standard
error for redshift bins, marked by dashed vertical lines, are shown in Table
\ref{tab-z_corr_alphaMedian}.}
	\label{fig:alphaZ8to20}
\end{figure}

As noted in Section 2, 838 out of 1460 of AT20G sources with redshift have 4.8 and 8.6 GHz
flux densities, 1 GHz NVSS/SUMSS counterparts and 20 GHz flux densities as well
as 6km visibility. We use this subsample of
sources for the following analysis. Of these, 683 sources are compact
by our
definition while the remaining 155 sources are extended. The distribution of
the redshifts for this AT20G subpopulation is listed in Table
\ref{tab-z_corr_alphaMedian}.

Fig. \ref{fig:alphaZ8to20} plots the spectral indices for
compact and extended sources between 8.6 and 20 GHz against redshift. It is
evident from the plot that extended sources do not show apparent changes in the
spectral index over the redshift range.  Compact sources, however, show a
distinct steepening of their spectra with redshift. It is apparent that a
spectral classification is no longer separating the compact and extended sources
above redshifts $\sim$ 1. 

Since most extended radio sources have close to power-law spectra
over a large range of radio frequencies, we do not expect a significant change
in spectrum with redshift due to the change in restframe frequency. But as
discussed in section 5, compact sources have flat spectra at lower frequencies
that gets steeper at higher frequencies. Hence, the spectral indices of compact
sources are affected by the redshift effect causing the redshift cut-off as
discussed by \citet{Jarvis2000}. To test this idea further we use the spectra at
different observed frequencies for different redshift ranges.  

\begin{figure}
	\centering
	\begin{center}
	\includegraphics[width=0.5\textwidth]{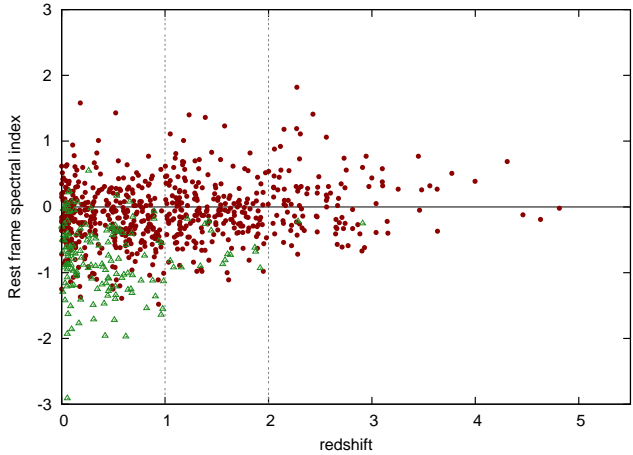}
\end{center}

	\caption{Rest frame spectral index of compact and extended sources.
Filled circles represent compact sources and open triangles represent
extended sources. All points have been brought close to the rest frame spectral index between 8.6 and 20 GHz by using the spectral index between 8.6 and 20 GHz for all sources below z = 1, using spectral indices between 4.8 and 8.6 GHz for all sources between redshifts 1 and 2 and
using spectral index between 1 and 4.8 GHz for all sources with z $\geq$ 2.}
	\label{fig:zDepAlphaCombined}
\end{figure}

The median spectral indices and errors for these redshift ranges are listed in
Table \ref{tab-z_corr_alphaMedian}. Our sources have flux densities measured
between four to five frequencies which allows us to calculate, at least, three
contiguous spectral indices ($\alpha_1^{4.8}$, $\alpha_{4.8}^{8.6}$,
$\alpha_{8.6}^{20}$).
We take advantage of these spectral points by using the best
representative spectra for different redshift bins to remove the effect of
redshift and, thus, compare their spectral indices between corresponding rest
frequencies. For this, we approximate that the spectral index of
$\alpha_1^{4.8}$ best represents the redshift bin between 2 and 5 (rest
frequencies $\sim$8 and $\sim$14 GHz), $\alpha_{4.8}^{8.6}$ best represents the
redshift bin between 1 and 2 (rest frequencies $\sim$10 and $\sim$26 GHz) and
$\alpha_{8.6}^{20}$ represents the redshift bin for redshifts up to 1. Plotted
in Fig. \ref{fig:zDepAlphaCombined}, this essentially brings all spectral
indices to $\alpha_{8.6}^{20}$ for comparison in their restframes. We use this
simple procedure to preserve the measured values rather than fitting complex
spectral shapes to a small number of frequencies.

\begin{table*}
%\begin{minipage}{126mm}

\caption{Median spectral indices and standard error for compact and extended
sources before (Fig. \ref{fig:alphaZ8to20}) and after (Fig.
\ref{fig:zDepAlphaCombined}) redshift corrections. There are only two extended
sources in the z $\geq$ 2 bin. The median and standard error are kept inside
parentheses to note their limited significance.}	
  \begin{normalsize}
  \begin{center}
  \begin{tabular}{lllllllllllll}
  %\begin{table}
	  \hline
	  Redshift& \multicolumn{5}{c}{Compact}
&&\multicolumn{5}{c}{Extended}&\\
	  \cline{2-6} \cline{8-12}
	   Bins 	& \textit{Number} & Uncorrected	& Error
& Corrected & Error && \textit{Number} &
Uncorrected &	Error	& Corrected	& Error & \textit{Total}\\
	  %\hline
	  \hline
	  z $<$ 1 	&\textit{363} & -0.17	&  0.02	& -0.17
& 0.02 &&\textit{138}& -0.79 & 0.05& -0.79	& 0.05 &
\textit{501}\\
	  1 $\leq$ z $<$2&\textit{211}& -0.28	&0.03& -0.07&
0.03 &&\textit{15}& -0.80 &0.08& -0.73&0.07 & \textit{226}\\
	  z $\geq$ 2 	&\textit{109}& -0.49&0.04	& 0.10&0.05
&&\textit{2}& (-0.57) &(0.38)&(-0.24)&(0.02)	 & \textit{111}\\
	  \hline
	  \textit{Total}	& \textit{683} & & & & &&\textit{155}&
& & & & \textit{838}\\
	  \hline

  \end{tabular}
  \end{center}
  \end{normalsize}
\label{tab-z_corr_alphaMedian}
%\end{minipage}
\end{table*}

\begin{figure}
	\centering
	\begin{center}
	\includegraphics[width=0.5\textwidth]{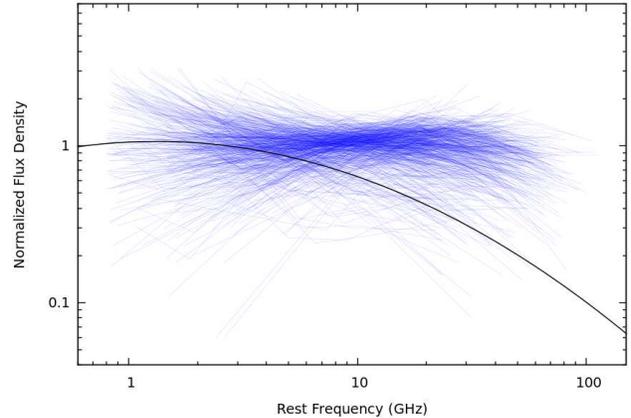}

\end{center}

	\caption{Rest frame radio spectra of 671 compact AT20G
sources. Sources have flux density observations at a subset
of 0.8, 1.4 and 4.8, 8.6 and 20 GHz. All observed frequencies are converted to the rest frame and the flux densities are normalized by their average
flux density over all frequencies present. The solid black line is the fit to
the spectra of the most luminous Parkes Half-Jansky Flat-Spectrum sample done by
\citet{Jarvis2000} normalized at 2.7 GHz survey frequency. }
	\label{fig:moustacheCompact}
\end{figure}

As is evident in Fig. \ref{fig:zDepAlphaCombined}, the redshift
correction completely removes the steepening of the spectral indices at higher
redshift seen in Fig. \ref{fig:alphaZ8to20}. Table
\ref{tab-z_corr_alphaMedian} even indicates that there is now an overall
flattening of spectra at higher redshift to 9 sigma confidence. These results
show that the spectral curvature plays a
dominant role at higher frequencies. To illustrate this further, we have
constructed the K-corrected spectra for the entire sample of sources in Fig.
\ref{fig:moustacheCompact}. Of the sub population of 1377 sources (see Table
\ref{tab-at20g_subpop}), 1092
sources are compact by our definition and only 671 sources have spectra at
four to five frequencies. We calculated the rest frequencies for
these sources and normalized them with their average flux density, and plotted
them against the rest frequency in Fig.
\ref{fig:moustacheCompact}. The mean redshift for these 671 compact sources
is $\sim$1. The figure confirms the results seen in fig. 6 of
\citet{Wall2005} which shows that the spectra of
flat-spectrum sources at their rest frames at lower frequencies is predominantly
flat suggesting that spectral curvature does not play a significant role at
lower frequencies.  At lower frequencies up to
$\sim$ 5 GHz, we observe that there is a mixture of steep, flat and inverted
spectrum sources. It is also clear from the figure that the majority of compact
sources remain flat-spectrum up to $\sim$30 GHz in their rest frame frequencies
at
all redshifts but the trend is then for almost all sources to get steeper.
The median normalized flux density for the rest frequency bin
between 35 - 45 GHz drops by $\sim$15$\%$ compared to that for the rest
frequency bin between 15 - 25 GHz. This
explains the effect of spectral steepening at
high frequency of compact, and hence, flat-spectrum sources at higher redshifts
seen in Fig. \ref{fig:alphaZ8to20} and Table \ref{tab-z_corr_alphaMedian}, as
well as the removal of the effect after redshift correction in Fig.
\ref{fig:zDepAlphaCombined}. The plot also indicates that at lower frequencies,
the spectral curvature does not play a significant role. In fact, there is a
population of steep-spectrum core-jet sources at
the lower frequency end of Fig. \ref{fig:moustacheCompact} which produces the
opposite effect. As discussed in Section 5, these compact sources that have
steep-spectrum jets at low
frequencies are increasingly dominated by their flat-spectrum cores at higher
frequencies. The redshift effect of these sources, then, makes them appear flat
spectrum at observed lower frequencies at higher redshifts. For sources at high
redshifts, this effect negates the effect of spectral steepening of the overall
spectra, at lower frequencies.

Thus, the redshift cut-off is
affected by the spectral curvature of compact sources as suggested by
\citet{Jarvis2001}, but this effect only becomes important at higher
frequencies ($\nu > $5 GHz) and can be avoided by measuring the spectra at
lower frequencies. Fortunately, this is where majority of evolution studies
have been done (e.g. \citet{Peacock1985, Dunlop1990,
Shaver1999, Jarvis2000, Wall2005}).

So far, we have only discussed the effect of the spectral curvature on the
redshift cut-off caused by the use of the spectral index to filter out the
compact flat-spectrum sources. However, it is also necessary to see how this
effects the number of sources above the flux limit of the survey, referred to
by \citet{Jarvis2000} as the observable volume. This cut-off
is determined by the survey frequency regardless of the frequency used to
measure the spectrum. Low frequency surveys will select against the compact
flat-spectrum sources at all redshifts. Higher frequency surveys will have
much larger populations of compact sources but they will also have a much
stronger redshift cut-off due to spectral curvature.

To evaluate this survey sensitivity effect, we plot the average parameters
used in
the \citet{Jarvis2000} models, normalized at 2.7 GHz, with our rest frame
spectra
in Fig.
\ref{fig:moustacheCompact}. The model is given by a polynomial of the form $ y
= $ $\rm{log_{10} S_{\nu}}$  = $\Sigma_{i=0}^2 a_i x^i$, where $x$  = 
$\rm{log_{10}}$($\nu$/GHz)  
%log_{10} S_\nu = a_0 + a_1 * log_{10} \nu + a_2 * (log_{10}\nu)^2 $ 
\rm{with}~ $a_1$ = 0.07 $\pm$ 0.08 and $a_2$ = -0.29 $\pm$ 0.06. 

It is apparent that the model predicts a much stronger spectral curvature than
that obtained for the AT20G compact sources. Even the closest fit, at the
extreme of the parameters range, is steeper than that obtained for the
AT20G
compact sources. We note that the data available to \citet{Jarvis2000} was
strongly biased to unusual
spectra and GPS sources by the spectral data available and is not representative
of the true compact source population. On the basis of their model spectrum
\citet{Jarvis2000} predicted a sharp cut-off in the number of observable
sources at high redshift and consequently argued that the cut-off seen by
\citet{Shaver1996} was not evidence for cosmological evolution. On average, our
spectra are at least five times brighter than \citet{Jarvis2000}
model, at our higher frequencies which are relevant for the high redshift
sources found in lower frequency samples such as those of \citet{Jarvis2000} or
\citet{Wall2005}. So, even at high redshift the expected number of sources
will be larger
by a similar fraction at this flux density level. The bias
introduced by preferential selection of flat-spectrum sources due to high AT20G
survey frequency is appropriate for the high redshift sources where cut-off is
seen. Hence, the cut-off reported by \citet[]{Shaver1996} must be mainly due to cosmological evolution
and is not the effect of K-correction on either the number of sources or the
ability to make spectral classification. The more detailed analysis of \citet{Wall2005}, 
which incorporated the various selection effects and biases explicitly, also demonstrated 
the reality of the cosmological evolution component.

We compared the Parkes quarter-Janksy flat-spectrum sample \citep{Jackson2002} with
the AT20G data in the overlapping area of the sky. The PQJFS sample has 804 southern hemisphere
sources and actually includes sources with flux densities less than 250 mJy. We have included all
sources down to 220 mJy since that corresponds to 100 mJy at 20 GHz for the spectral index cut-off
of \mbox{-0.4}.
There are 716 PQJFSS sources over the limit of 220 mJy at 2.7 GHz. The cross match of these
sources' positions against the AT20G catalogue \citep{Murphy2010} finds 640 of these sources and
further 31 are found in the AT20G scanning survey catalogue \citep{Hancock2011} which
avoids the incompleteness introduced by the AT20G follow-up observations. We used the
updated positions of the sources provided in \citet{Jackson2002}, where available, since the
original PQJFSS positions are not fully reliable. Thus, our
cross-match found a total of 671 (94 $\%$) PQJFSS sources in the AT20G survey. Of the remaining 45
sources, 14 were later found to be extended sources by \citet{Jackson2002} and are not included in
their
analysis. For the remainder, our literature search finds that 7 are detected at lower frequencies
and have steep-spectrum, 9 are known variable sources based on multiple epoch observations and 1
source that has not been detected in any other surveys. The remaining 14 sources (2$\%$) are well
within the completeness expected from long term variability noting that the AT20G survey was done
$>$ 20 years after the PQJFS survey. Only 3 (0.4$\%$) of these sources have evidence for extreme
spectral steepening at high
frequency. Hence, we can conclude that we have not lost a significant fraction of sources as a
result of spectral steepening which has put them below our survey limit.

603 of the detected sources have 6km visibilities and 548 of them are compact while 55 (9$\%$) are
extended, consistent with the overlap in the spectral index distribution shown in Fig.
\ref{fig:histo_alpha-6kvis}. Out of the detected sources, 461 sources have redshift information.
The published PQJFS sample has 459 sources in the southern hemisphere with redshifts.

\section{Summary}
We summarise the results of this work.

\begin{enumerate}

	\item 6km visibility, calculated as the ratio of five long
baselines to the 10 short baselines of the ATCA in hybrid configurations,
provides information on the angular sizes of the extragalactic source
population at 20 GHz. It is used to separate the source
population into compact and extended sources at subarcsec (0.15 arcsecond)
level.
	\item 6km visibility provides a firm physical basis for spectral
classification into flat and steep-spectrum sources to select compact and
extended sources. For spectral indices measured between 1 and 5 GHz,
the spectral index of -0.46 is the optimum cut-off between the two populations.
However, the traditional cut-off of either -0.4 or -0.5 are almost as good and
for consistency, we recommend that -0.5 be used in future studies.
	\item 79\% of the sources in our
survey are compact AGNs which is a result of the high survey frequency. This is
very different from low frequency surveys such as NVSS which has only 25\% of
the flat-spectrum AGN source population.
	\item A larger scatter in the spectral index of extended sources is
observed at higher frequencies due to the domination of compact cores in
core-jet sources as well as the steepening of extended jets due to energy loss.
Hence, the use of spectral index at high frequencies is not as effective to
select compact and extended sources. Using the radio spectrum below a few
gigahertz avoids this problem. 
	\item The spectral curvature is significant for compact
sources, but it only becomes significant at higher frequencies. Below a few
gigahertz frequencies, the redshift effect on
compact sources is not significant. We note that the general spectral
steepening seen above 20 GHz will occur at observed frequencies around 5 GHz,
for sources with redshift $>$ 3. Hence, surveys at frequencies $>$ 5 GHz will
have
a redshift cut-off due to spectral curvature.
	\item At high frequencies, the spectra of both compact and extended
sources are steep and the two populations become indistinguishable from
each other due to the spectral steepening of compact sources by $\sim$ 15\%
between 20 and 40 GHz. 
	\item The effect of K-correction shown by \citet{Jarvis2000} is
important at high frequencies as discussed by \citet{Shaver1996} and
\citet{Wall2005} for flat-spectrum sources and also for steep-spectrum sources
\citep{Rigby2011}. However, this should not affect the results of
\citet{Shaver1996} and \citet{Wall2005} as they used low frequency spectral
index to select compact sources. Hence, the conclusion that the compact sources
show a strong redshift cut-off is supported by our data.

In future, with the information on the completeness of the optical
identification and redshifts, it will be possible to explore the density
evolution and redshift cut-off of compact AGN without using the spectral index
to select compact sources.

\end{enumerate}

\section*{Acknowledgements}
We thank the staff at the ATCA site, Narrabri (NSW), for the valuable support
they provided in running the telescope. The ATCA is part of the Australia
Telescope which is funded by the Commonwealth of Australia for operation as a
National Facility managed by the CSIRO.

R.C. would like to thank Tara Murphy for providing cross-matching scripts. MM
acknowledges financial support for this research by ASI (ASI/INAF Agreement
I/072/09/0 for the Planck LFI activity of Phase E2 and contract I/016/07/0
COFIS).

\bibliographystyle{scemnras}
%\bibliography{mn2e}
\bibliography{references}

\bigskip
This document has been typeset from a \TeX/\LaTeX  file prepared by the
author.
\end{document}